\documentclass[twocolumn,nopacs,floatfix,amsmath,nofootinbib,amssymb,preprintnumbers]{revtex4-1}
\usepackage{graphicx,color,dcolumn,booktabs,bm}
\usepackage{txfonts}
\usepackage{slashed}
\usepackage[colorlinks,
            citecolor=blue,
            anchorcolor=red,
            menucolor=red,
            linkcolor=red,
            filecolor=red,
            runcolor=red,
            urlcolor=blue,
            frenchlinks=red]{hyperref}

\begin{document}

\title{Potential of the reaction $e^+e^-\to p\bar{p}\pi^0$ for constructing higher $\rho$-meson spectroscopy above 2.4 GeV} 

\author{Dan Guo}
\affiliation{School of Physics and State Key Laboratory of Nuclear Physics and Technology, Peking University, Beijing 100871, China}
\author{Jun-Zhang Wang}\email{wangjzh@cqu.edu.cn}
\affiliation{Department of Physics, Chongqing University,
Chongqing 401331, China}
\affiliation{School of Physics and Center of High Energy Physics, Peking University, Beijing 100871, China}
\author{Qin-Song Zhou}\email{zhouqs@imu.edu.cn}
\affiliation{School of Physical Science and Technology, Inner Mongolia University, Hohhot 010021, China}
\affiliation{Research Center for Quantum Physics and Technologies, Inner Mongolia University, Hohhot 010021, China}
\affiliation{Inner Mongolia Key Laboratory of Microscale Physics and Atomic Manufacturing, Hohhot 010021, China}

\begin{abstract}
The spectrum of light unflavored vector mesons above 2.4 GeV has not yet been firmly established experimentally. In this work, we demonstrate that the process $e^+e^- \to p\bar{p}\pi$ serves as a particularly promising channel as a probe for higher isovector $\rho$ excitations. In contrast to typical electron-positron annihilation processes into purely light-meson final states, the reaction $e^+e^- \to p\bar{p}\pi$ benefits from a relatively low continuum background, on the order of $(50\sim100)~\text{pb}$ around 2.4 GeV, which enhances the visibility of even modest resonance contributions. In addition, the comparatively high production threshold further suppresses potential contamination from lower $\rho$-meson excitations. Our analysis based on effective Lagrangian approach indicates that the observed lineshape of the $e^+e^- \to p\bar{p}\pi$ cross section measured by BESIII can be naturally interpreted by including the $\rho(5S)$ and $\rho(6S)$ contributions, namely the fourth and fifth radial excitations of $\rho(770)$. In particular, the pronounced rise of the cross section near 2.4 GeV provides strong evidence for the contribution of $\rho(5S)$ signal. Their branching ratios to the $p\bar{p}\pi$ final state are found to be below 4\%, which do not contradict with usual expectations. In addition, we evaluate pion-induced production of these two $\rho$-meson states as a complementary probe, offering useful guidance for the relevant experiments such as COMPASS.
\end{abstract}

\maketitle
\section{Introduction}\label{sec:intro}

Light hadron spectroscopy provides a rich laboratory for exploring the strong interaction in the nonperturbative regime, where the color confinement dynamics are further complicated by active sea-quark effects inside hadrons and relativistic contributions become significant. Moreover, many candidates for exotic hadronic states such as glueballs or hybrids are expected to emerge in this sector~\cite{Klempt:2007cp,Ochs:2013gi}. Within the light hadron family, unflavored vector mesons are of particular interest because their production can be directly accessed in electron–positron annihilation experiments such as \textit{BABAR}~\cite{BaBar:2014omp}, Belle~\cite{Belle-II:2018jsg} and BES~\cite{Asner:2008nq,BESIII:2020nme}, which have accumulated extensive and high-precision datasets over the past decades~\cite{Godfrey:1998pd,Fang:2020aqg,Huang:2020dga}. 

The spectroscopy of  highly excited mesons usually offer a sensitive probe of the long-distance behaviors of color confinement, making systematic investigations of the $\rho$-, $\omega$-, and $\phi$-like excitations an important task for hadron physics. Research on highly excited states of light unflavored vector mesons has so far reached up to approximately 2.2 GeV. In the $\phi$-meson family, the most well-known example is $\phi(2175)$~\cite{ParticleDataGroup:2024cfk,BaBar:2006gsq}, initially proposed as a strange-sector partner of the charmonium-like state $Y(4260)$~\cite{BaBar:2005hhc,Zhu:2005hp}. This state has attracted extensive theoretical discussions, and one popular interpretation is strangemonium excitation $\phi(3S)$ or $\phi(2D)$~\cite{Barnes:2002mu,Pang:2019ttv,Wang:2012wa, Ding:2007pc}. A theoretical combined analysis of electron-positron annihilation to open-strange channels shows that the $\phi(2175)$ signal should simultaneously involve contributions from both the vector states $\phi(3S)$ and $\phi(2D)$, whose theoretical masses and decay behaviors based on an unquenched relativized potential model are taken as input of the fit analysis~\cite{Wang:2021gle}. For the $\omega$-meson family, the BESIII Collaboration recently had observed two enhancement structures near 2.2 GeV existing in $e^+e^- \to \omega\eta$ and $e^+e^- \to \omega\pi^0\pi^0$~\cite{BESIII:2020xmw,BESIII:2021uni}, which have been suggested as the interference structures from highly excited states $\omega(4S)$ and $\omega(3D)$~\cite{Zhou:2022wwk}. In addition, the authors in Ref.~\cite{Bai:2025knk} introduced the $4S$-$3D$ mixing effect in the highly excited $\omega$ spectroscopy to explain the $\omega$-like structure $Y(2119)$ recently observed in $e^+e^- \to \rho\pi$ and $e^+e^- \to \rho(1450)\pi$~\cite{BESIII:2024okl}.

\begin{figure*}[!tpb]
  \centering 
  \begin{tabular}{rr}
    \includegraphics[width=0.4\textwidth]{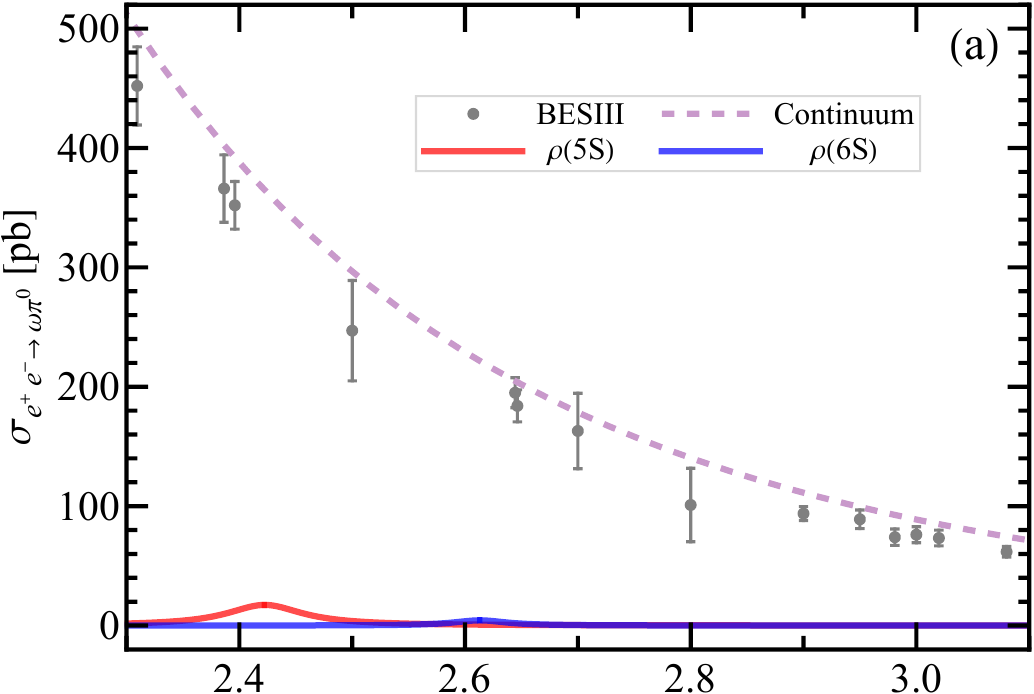} & \includegraphics[width=0.4\textwidth]{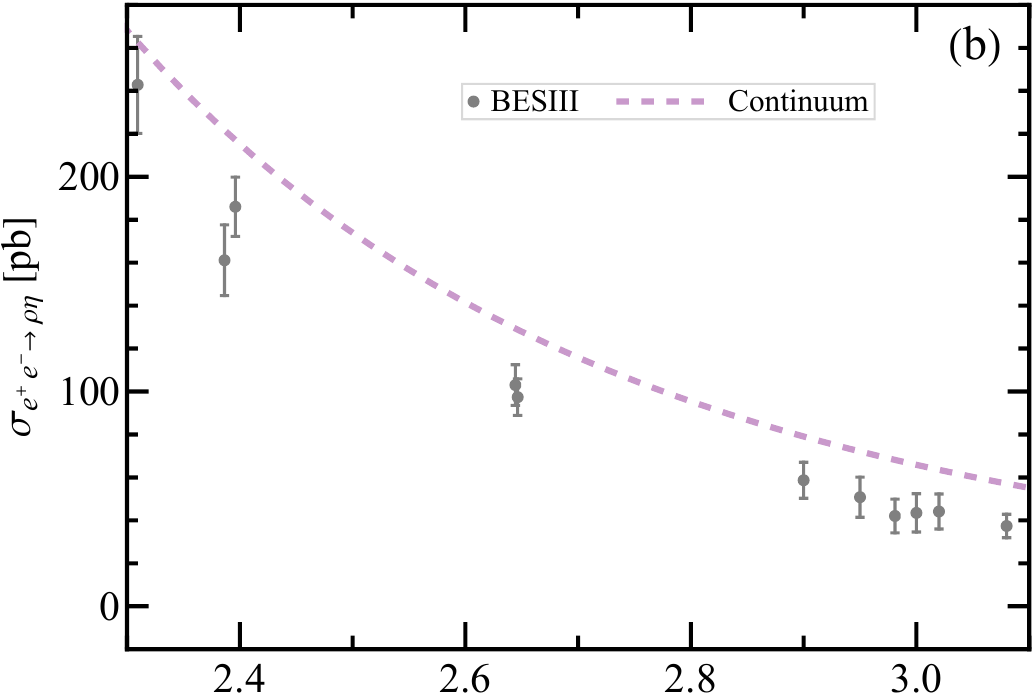} \\
    \includegraphics[width=0.4\textwidth]{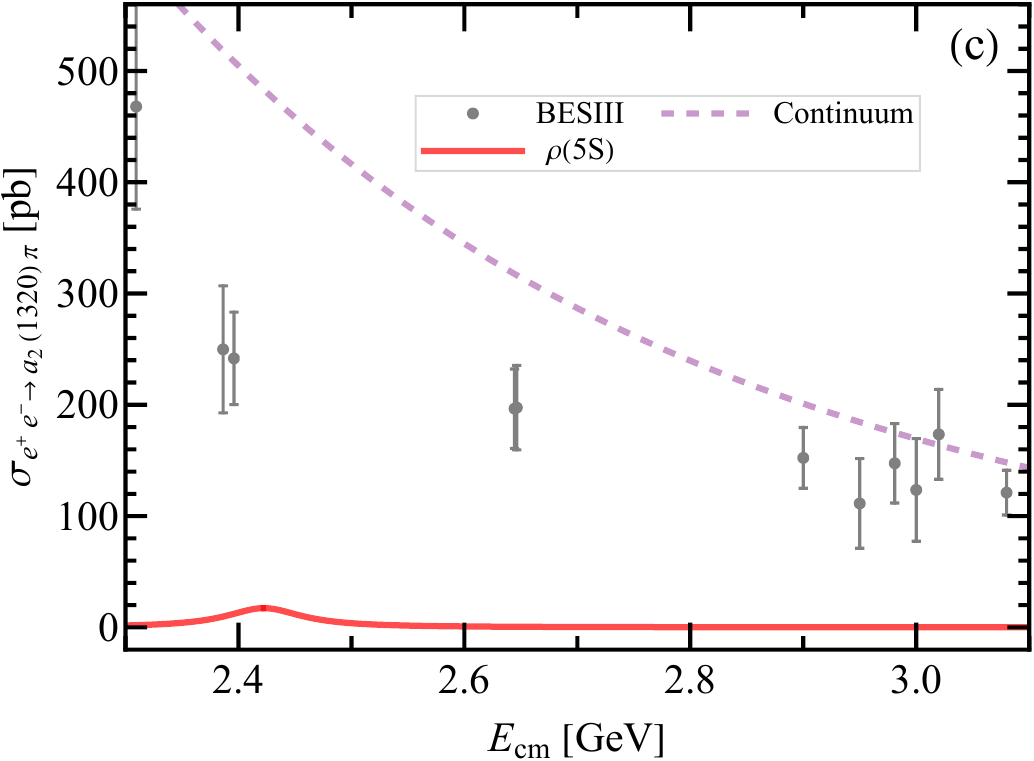} & \includegraphics[width=0.4\textwidth]{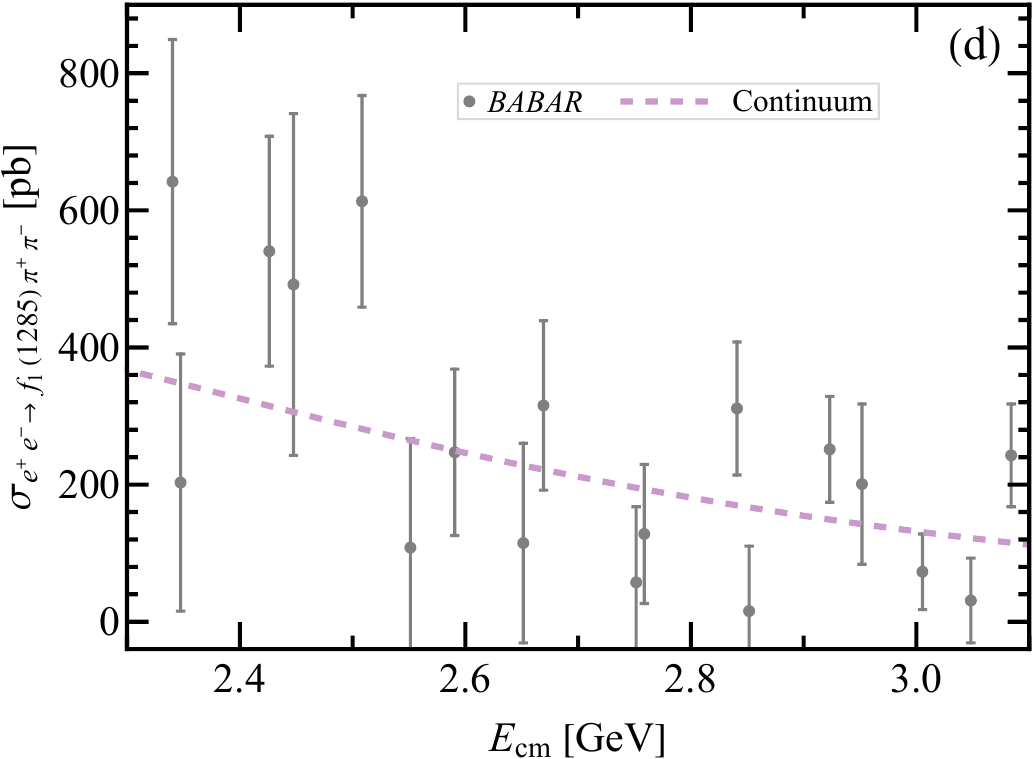} \\
  \end{tabular}
  \caption{Total cross sections of (a)~$e^+e^-\to \omega\pi^0$~\cite{Achasov:2016zvn,BESIII:2020xmw}, (b)~$e^+e^-\to \rho \eta$~\cite{BESIII:2023sbq}, (c)~$e^+e^-\to a_2(1320)\pi^0$~\cite{BESIII:2023sbq}, (d)~$e^+e^-\to f_1(1285)\pi^+\pi^-$~\cite{BaBar:2007qju,BaBar:2022ahi} above 2.3 GeV. The fitted continuum and theoretical resonance contributions of $\rho(5S)$ and $\rho(6S)$ are from Ref.~\cite{Zhou:2025kmw}. Here, the reason for the absence of the resonance contribution in subfigure (b) and (d) is the tiny decay branching ratio of $\rho(5/6S)$ to the corresponding final states. }\label{fig:diff4in1}
\end{figure*}

In the isovector $\rho$-meson family, there are three excited states $\rho(1900)$, $\rho(2000)$ and $\rho(2150)$ around 2 GeV collected in Particle Data Group (PDG) \cite{ParticleDataGroup:2024cfk}, whose dominant components were usually considered as $\rho$ meson excitations $\rho(3S)$, $\rho(2D)$ and $\rho(4S)$, respectively, in various theoretical approaches such as potential model \cite{Godfrey:1985xj,Barnes:1996ff,Ebert:2009ub,Li:2021qgz}, Regge trajectory \cite{He:2013ttg,Feng:2021igh} and other phenomenological models \cite{Hilger:2015ora,Branz:2010ub,Yu:2021ggd}. In recent years, the BESIII Collaboration performed the precise measurement of the cross sections for several isospin vector processes and also observed some $\rho$-like enhancement structures around 2 GeV~\cite{BESIII:2020xmw,BESIII:2023sbq,BESIII:2020kpr}. According to a recent theoretical analysis for the six electron-positron annihilation processes with different light meson final states, the line shapes of these enhancement structures can also be well reproduced by introducing the contribution from higher $\rho$ meson excitations around 2.2 GeV~\cite{Zhou:2025kmw}.

Despite these advances, the higher light vector meson states above 2.2 GeV  have remained elusive experimentally. The search for such higher excitations is undoubtedly crucial for extending our understanding of the spectroscopy of light vector mesons, and is essential for testing the validity of various theoretical models. In Refs. \cite{Wang:2021abg}, an unquenched relativized potential model was employed to predict the next $\rho$-meson excitations, namely the $\rho(5S)$ and $\rho(6S)$, should lie around 2.4 GeV and 2.6 GeV, respectively, both with typical widths of about 100 MeV. Their isoscalar partners in the $\omega$-meson family are also expected in this mass region, but none of these states have yet been discovered experimentally. Traditionally, searches for higher $\rho$ excitations above 2.4 GeV should focus on typical light-meson final states, as observed for the $\rho$-meson states near 2 GeV, since such channels are generally considered to be their dominant decay modes. In contrast, baryonic final states, such as those involving a nucleon–antinucleon pair, are usually disfavored in this context, because the decay into these channels requires the creation of an additional light quark pairs from the vacuum, which suppresses their branching ratios. As a result, these modes have often been overlooked in previous searches.

In this work, however, we point out that the process $e^+e^-\to p\bar{p}\pi$ provides a promising alternative channel for exploring the $\rho(5S)$ and $\rho(6S)$ states. Recently, BESIII reported precise measurements of the $e^+e^- \to p\bar{p}\pi$ cross section~\cite{BESIII:2024gjj}. Although the available energy points are not sufficiently dense, we notice that a nontrivial rapidly enhanced point around 2.4 GeV, intriguingly close to the predicted mass of the $\rho(5S)$, seems difficult to be produced by the continuum term. Motivated by this, we perform a detailed analysis of the cross section using an effective Lagrangian approach,  and find strong evidence suggesting the presence of both the $\rho(5S)$ and $\rho(6S)$ states. It is worth noting that the role of the $\omega(5S)$ and $\omega(6S)$ in this process basically can be ruled out. Moreover, two different analysis schemes consistently show that the branching fractions of $\rho(5S)$ and $\rho(6S)$ decays into the $p\bar{p}\pi$ final state do not exceed $4\%$, which is consistent with the expected suppression mechanism.

Although the $p\bar{p}\pi$ mode is not expected to be the dominant decay channel, it is important to emphasize that the visibility of a resonance signal in experiment does not solely depend on its coupling strength to a given final state. Instead, the relative size of the resonance contribution with respect to the continuum background plays a decisive role. If the continuum background is overwhelmingly large, even a signal with a sizable branching ratio may still be obscured. In the case of $e^+e^- \to p\bar{p}\pi$, a key advantage lies in its relatively small background contribution, at the level of about $(50\sim100)$ pb around 2.4 GeV~\cite{BESIII:2024gjj}, which leaves room for the $\rho(5S)$ and $\rho(6S)$ signals to emerge.
For comparison, in  Fig.~\ref{fig:diff4in1} we present the cross section data between 2.3 and 3.1 GeV for the $e^+e^-$ annihilation channels into different light mesons together with the theoretical resonance contributions from the $\rho(5S)$ and $\rho(6S)$ states~\cite{Zhou:2025kmw}, including $e^+e^-\to \omega\pi^0$~\cite{Achasov:2016zvn,BESIII:2020xmw}, $e^+e^-\to \rho \eta$~\cite{BESIII:2023sbq}, $e^+e^-\to a_2(1320)\pi^0$~\cite{BESIII:2023sbq} and $e^+e^-\to f_1(1285)\pi^+\pi^-$~\cite{BaBar:2007qju,BaBar:2022ahi}. In all these channels, the continuum backgrounds are much larger, which may partly explain why the two higher $\rho$-meson excitations are still missing. Another notable advantage of the $e^+e^- \to p\bar{p}\pi$ reaction is its relatively high threshold, which strongly suppresses the contributions from lower $\rho$-meson excitations. From an experimental perspective, the proton in the final state is more stable compared with purely light-meson final states, offering further benefits in data reconstruction.

Finally, motivated by these findings, we have also calculated the pion-induced production of the $\rho(5S)$ and $\rho(6S)$ states and predicted the corresponding cross section, including both contact and $t$-channel amplitudes. These results may provide useful input for future searches for the $\rho(5S)$ and $\rho(6S)$ states at pion-proton collision facilities such as COMPASS~\cite{Ketzer:2019wmd}.

This paper is organized as follows. After the Introduction, Sec.~II presents a general analysis of $t$-channel dominance in baryonic and mesonic production mechanisms. In Sec.~III, we investigate the contributions of the higher $\rho(5/6S)$ resonances to the $e^+e^-\to p\bar{p}\pi^0$ process, including both resonant and nonresonant amplitudes, and perform fits to the BESIII data. In Sec.~IV, we extend our study to pion-induced reactions $\pi p \to \rho(5/6S)p$, where both contact and $t$-channel exchange mechanisms are considered, and we provide predictions for branching cross sections into different final states. Finally, Sec.~V contains the summary.



\section{dominance analyses in the $\rho^* \to p\bar{p}\pi^0$ and $\pi p \to \rho^* p$ processes}\label{sec:domi_anal}

\begin{figure*}[thbp]
    \centering
    \includegraphics[width=\textwidth]{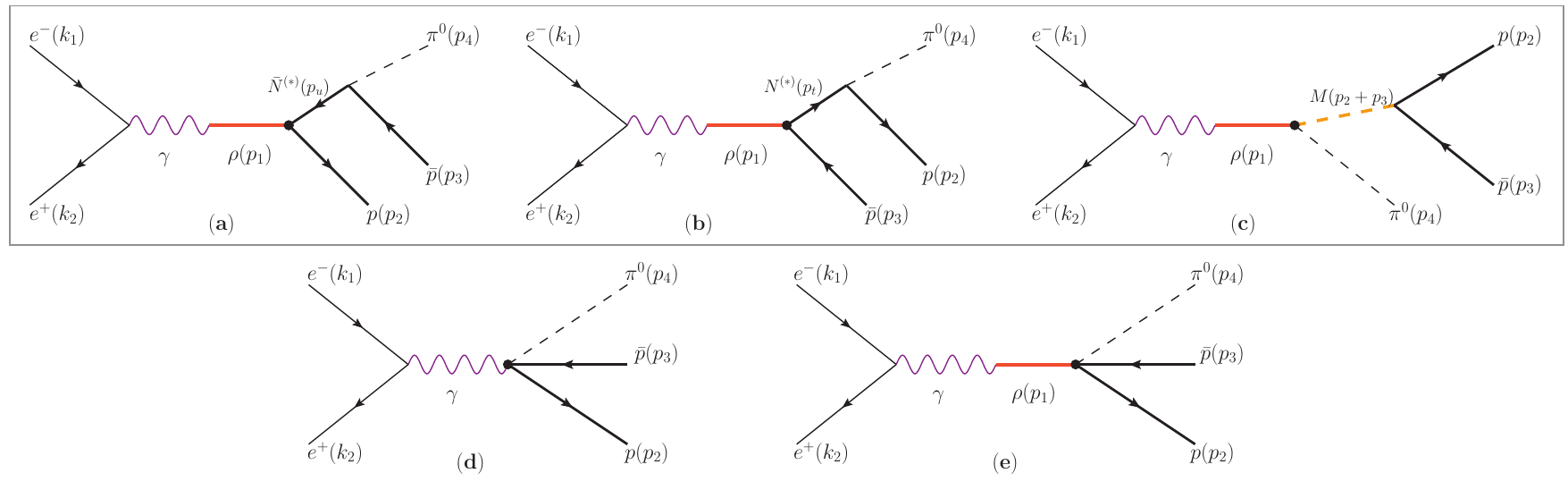}
    \caption{Feynman diagrams for the process $e^+e^-\to p\bar{p}\pi^0$. The first-row diagrams (a)–(c) denote the resonant contributions from the cascade decays of $\rho(5/6S)$, which are found to be negligible in present case. Diagrams (d) and (e) correspond to the direct background and the direct resonant contribution, respectively.}
    \label{Feynman1}
\end{figure*}

To clarify the production mechanisms in $e^+e^-\to p\bar{p}\pi^0$, we first analyze the relative importance of different topologies.  
For the resonant diagrams $e^+e^-\to \rho^*(p_1)\to (M\to p\bar{p}) (p_3,p_4) \pi(p_2)$ and 
$e^+e^-\to \rho^*(p_1)\to (N^*\to p\pi)(p_3,p_2)\bar{p}(p_4)$, in which $\rho^*$ stands for higher excited vector meson state, the corresponding amplitudes are denoted as 
\begin{align}
    \mathcal{M}_1&\propto \frac{1}{m_{34}^2-m_{c}^2}, \\
     \mathcal{M}_2&\propto \frac{1}{m_{23}^2-m_{c}^2}
\end{align}
with the kinematic constraint $m_3+m_4 < m_{34} < m_1 - m_2$ and $m_2+m_3 < m_{23} < m_1 - m_4$.  Here, $m_c$ is the mass of intermediate meson state $M$ or nucleon state $N^*$.
In the limit $m_\pi \simeq 0$ and $m_N\simeq 1$, the amplitudes approximately scale as
\begin{align}
    \mathcal{M}_1 &\propto \frac{1}{[4m_N^2,m_{\rho^*}^2]-m_{c}^2}\propto \frac{1}{\mathcal{O}(4m_{N}^2)-m_{c}^2}, \\
    \mathcal{M}_2 &\propto \frac{1}{[m_N^2,(m_{\rho^*}-m_N)^2]-m_{c}^2}\propto \frac{1}{\mathcal{O}(m_{N}^2)-m_{c}^2}.
\end{align}
This comparison indicates that we should retain three kinds of dominant mechanisms for resonant diagrams of $e^+e^-\to p\bar{p}\pi^0$  in subsequent analysis: (1)~the $\rho^*$ couples to $M \pi$ with a mass of intermediate meson $M$ around the $p\bar{p}$ threshold; (2)~the $\rho^*$ couples to $N^*\bar{p}$ with an intermediate nucleon $N^*=p$; (3)~the direct $p\bar{p}\pi$ coupling of the $\rho^*$ resonance. 

For meson-induced reactions such as $\pi p$ or $K p$ scattering to $\rho^*p$, the situation differs: both $t$-channel (light-meson exchange) and $u$-channel (nucleon exchange) diagrams contribute. In these cases, the generic amplitude takes the form
\begin{eqnarray}
 \mathcal{M}_1\propto \frac{1}{t/u-m_c^2}.
\end{eqnarray}
By setting $s = (m_N + m_{\rho^*} + \delta)^2$ with $m_\pi/m_K=\epsilon$, and expanding to leading order in $\mathcal{O}(\delta)$ and $\mathcal{O}(\epsilon)$, one obtains
\begin{align}
    t_{min}&=t_0-\Delta t,\quad t_{max}= t_0+\Delta t, \\
    u_{min}&=u_0-\Delta u,\quad u_{max}= u_0+\Delta u, 
\end{align}
where
\begin{align}
    t_0 =&-\frac{1}{(m_N+m_{\rho^*})^2}(m_N^2m_{\rho^*}^2+m_Nm_{\rho^*}^3 \nonumber \\ 
    &+\delta(m_{\rho^*}^3+4m_N^2m_{\rho^*}+5m_Nm_{\rho^*}^2)),   \nonumber \\
    \Delta t =&\frac{3m_Nm_{\rho^*}^2+m_{\rho^*}^3+2m_N^2m_{\rho^*}}{(m_N+m_{\rho^*})^3} \nonumber\\
    &\times \sqrt{\delta(2m_N^2m_{\rho^*}+2m_Nm_{\rho^*}^2)+\delta^2(m_N^2+3m_Nm_{\rho^*}+m_{\rho^*}^2)}, \nonumber\\
     u_0 =&-\frac{1}{(m_N+m_{\rho^*})^2}(-m_N^4+2m_N^2m_{\rho^*}^2+m_Nm_{\rho^*}^3 \nonumber \\ 
     &+\delta(m_{\rho^*}^3+4m_N^2m_{\rho^*}+5m_Nm_{\rho^*}^2)),   \nonumber \\
       \Delta u =&\frac{3m_Nm_{\rho^*}^2+m_{\rho^*}^3+2m_N^2m_{\rho^*}}{(m_N+m_{\rho^*})^3} \nonumber\\
    &\times \sqrt{\delta(2m_N^2m_{\rho^*}+2m_Nm_{\rho^*}^2)+\delta^2(m_N^2+3m_Nm_{\rho^*}+m_{\rho^*}^2)}. \nonumber\\
\end{align}
Since $\Delta u = \Delta t > 0$ and both $t_0$ and $u_0$ are negative, their magnitudes can be directly compared. It can be seen that $t_0$ or $u_0$ consists of a constant term and a term proportional to variable $\delta$, relating to the center-of-mass energy $\sqrt{s}$, while the $\Delta t$ term is entirely linear in $\delta$. Near the production threshold, $\delta$ should be a small quantity, so the $t/u$ always takes negative values. Consequently, one can conclude that the smaller the mass of the exchanged particle, the larger its contribution through the $t$- or $u$-channel near production threshold. To further compare the relative importance of the $u$- and $t$-channel contributions, we define the relative discrepancy as
\begin{eqnarray}
    \mathcal{A}=\frac{(u_0-t_0)}{u_0},
\end{eqnarray}
one finds
\begin{equation}
    \mathcal{A}_{max}=\mathcal{A}|_{\delta=0}=\frac{m_N(m_N -m_{\rho^*})}{m_N^2 -m_N m_{\rho^*} -m_{\rho^*}^2}.
\end{equation}
Numerically, $\mathcal{A}_{\max}=0.192$ for $\rho(5S)$ and $\mathcal{A}_{\max}=0.187$ for $\rho(6S)$.  
These values indicate that the $t$- and $u$-channel contributions are comparable for exchange particles with equal mass.  
At lower energies, however, exchanges involving lighter mesons dominate in the $t$-channel, whereas the nucleon-exchange 
contribution in the $u$-channel is strongly suppressed. Consequently, in the following calculations  we primarily focus on the pion exchange in the $t$-channel together with the direct contact coupling for the $\pi p \to \rho^* p$ reaction.

\section{higher excited $\rho(5/6S)$ resonance contributions in $e^+e^-\to p\bar{p}\pi^0$}\label{sec:fit}


In $e^+e^-$ collisions, the annihilation of an electron and a positron produces a virtual photon that subsequently couples to the final-state particles or other intermediate particles. Within the effective Lagrangian method, the process $e^+(k_2)e^-(k_1) \to p(p_2)\bar{p}(p_3)\pi^0(p_4)$ proceeds through both resonant and nonresonant channels.

In the energy region around $2.5~\mathrm{GeV}$, the modified Godfrey–Isgur model predicted the existence of unflavored excited light vector mesons belonging to the $\rho(5/6S)$ and $\omega(5/6S)$~\cite{Wang:2021gle,Wang:2021abg}. This prediction is obtained by successfully reproducing the mass spectrum of the low-lying light mesons. Since the dielectron widths of the $\omega$-meson excitations  are usually found to be about one order of magnitude smaller than those of the corresponding $\rho$-meson partners~\cite{Wang:2021gle,Wang:2021abg}, we therefore give priority to the resonance contributions from the $\rho(5S)$ and $\rho(6S)$ states in our analysis. The relevant resonance parameters, including the predicted masses, total widths, dilepton widths, and partial decay branching ratios, are collected in Table~\ref{tab:branch}, which provide the essential input for the present analysis. The resonant contributions include the decays of the intermediate $\rho(5/6S)$ resonance state. 
These decay includes:
\begin{itemize}
    \item $\rho(5/6S) \to N\bar{N}^*$ (Fig.~\ref{Feynman1} (a)) and $\rho(5/6S) \to \bar{N}N^*$ (Fig.~\ref{Feynman1} (b)), where the two processes are distinguished by the momentum topologies of the intermediate nucleons or antinucleons, denoted as $p_t$ and $p_u$.
    \item $\rho(5/6S) \to \pi^0 M$ (Fig.~\ref{Feynman1} (c)), where $M$ denotes a possible intermediate meson such as $\pi$, $\rho$, $\omega$, or $a_2(1320)$, among others.
    \item the direct three-body decay $\rho(5/6S) \to p\bar{p}\pi^0$ (Fig.~\ref{Feynman1} (e)), which belongs to the contact term.
\end{itemize}
In addition, the nonresonant contribution corresponds to a direct coupling with virtual photon (Fig. \ref{Feynman1} (d)).

In principle, all five diagrams shown in Fig.~\ref{Feynman1} should be considered, and the intermediate $N^{(*)}$, $\bar{N}^*$ and $M$ all corresponding to several possible candidate resonance in the relevant energy region. However, as discussed in Sec.~\ref{sec:domi_anal}, the dominant nucleon and meson resonances are expected to have masses near the $p\bar{p}$ and $\pi N$ thresholds, respectively.
Based on model predictions on light meson decays into baryon-antibaryon pair~\cite{Xiao:2019qhl,Bai:2023dhc}, the branching fraction of $\rho(5/6S) \to p\bar{p}$ can be roughly estimated to be at the level of $10^{-3}$. Based on this, we calculate the peak cross sections of the subprocesses in Fig.~\ref{Feynman1} (a) and (b) as around $0.06~\mathrm{pb}$, which is about three orders of magnitude smaller than the experimental measurements~\cite{BESIII:2024gjj}. 
From Table~\ref{tab:branch}, it can be seen that the $\pi^0M$ decay ratios with an intermediate meson $M$ near $p\bar{p}$ threshold are too small for both $\rho(5S)$ and $\rho(6S)$ that their numerical results are not provided in Ref.~\cite{Wang:2021abg}.  For the channels shown in Fig.~\ref{Feynman1} (c), taking $\rho(5S)\to (\rho(2150)\to p\bar{p})\pi^0$ as an example, the corresponding peak cross section is estimated to be about $8\times 10^{-3}~\mathrm{pb}$ if the branching ratio $\rho(5S)\to\pi\rho(2150)$ and $\rho(2150) \to p\bar{p}$ are taken to be at most 1\% and $10^{-3}$.
Therefore, the contributions from the intermediate states in Fig.~\ref{Feynman1} (a), (b), and (c) are negligible compared to the experimental data, and in our calculation we mainly focus on the direct background term (Fig.~\ref{Feynman1} (d)) and the dominant direct resonant contact term (Fig.~\ref{Feynman1} (e)). 


To account for the coupling between the photon and the vector mesons, we adopt the vector meson dominance (VMD) mechanism. Within this framework, the $\gamma\rho(5/6S)$ interaction is described by the effective Lagrangian
\begin{equation}
	\mathcal{L}_{\gamma V}=\frac{-eM_V^2}{f_V}V_{\mu}A^{\mu},
\end{equation}
where $M_V$ and $f_V$ denote the mass and decay constant of the vector meson, respectively, and $V_\mu$ and $A_\mu$ represent the vector meson and photon fields. The constant $f_V$ is determined by the dilepton decay width $\Gamma_{V\to e^+e^-}$ through the relation
\begin{equation}\label{}
  \frac e{f_V}=\left[\frac{3\Gamma_{V\to e^+e^-}M_V^2}{\alpha(M_V^2-4m_e^2)^{3/2}}\right]^{1/2},
\end{equation}
with the fine-structure constant $\alpha=1/137$. In Refs.~\cite{Wang:2021abg,Wang:2021gle}, the dilepton widths of light unflavored vector mesons were systematically calculated. 

\renewcommand{\arraystretch}{1.5}  
\begin{table}[htbp]
  \centering
  \caption{The masses, total widths, dilepton widths and decay branching ratios of $\rho(5/6S)$ calculated in Refs.~\cite{Wang:2021gle,Wang:2021abg}. A centerdots (...) denotes that branching ratios of $\rho(5/6S)$ are negligible. }\label{tab:branch}
  \setlength{\tabcolsep}{10pt} 
  \begin{tabular}{l|cc} 
  \toprule[1.50pt] \toprule[0.50pt]
         &$\rho(5S)$     &$\rho(6S)$ \\
  \hline
  $M_V$         &2422 MeV       &2613 MeV  \\
  $\Gamma_{total}$    & 81.85 MeV     & 57.25 MeV \\
  $\Gamma_{V\to e^+e^-}$ & 0.0377 keV    & 0.0204 keV \\ \hline
  $\pi a_2(1700)$   & 13.61\%  & 10.62\% \\
  $\rho\rho(1450)$  &  9.41\%  &  9.13\% \\
  $\pi\pi(1300)$    &  8.40\%  &  7.31\% \\
  $\pi\pi_2(1880)$  &  ...     &  7.56\% \\
  $\omega a_0(1450)$&  7.83\%  &  4.07\% \\
  $\pi\omega(1420)$ &  7.56\%  &  5.04\% \\
  $\pi\pi(1800)$    &  6.97\%  &  7.33\% \\
  $\pi a_2(1320)$   &  4.11\%  &  ... \\
  $\pi h_1(1170)$   &  3.28\%  &  ... \\
  $\pi\pi$          &  2.33\%  &  1.73\% \\
  $\pi\omega$       &  1.50\%  &  0.57\% \\
  \toprule[0.50pt] \toprule[1.50pt]
\end{tabular}
\end{table}
\setlength{\extrarowheight}{0pt}

For the direct background contribution shown in Fig.~\ref{Feynman1} (d), the virtual photon produced in $e^+e^-$ annihilation couples directly to the $p\bar{p}\pi^0$ final state without involving intermediate resonances. The corresponding amplitude can be expressed as~\cite{Xu:2015qqa,Wang:2017sxq,Chen:2010nv}
\begin{equation}\label{}
  \mathcal{M}_{\mathrm{NoR}}=g_{\mathrm{NoR}}\bar{v}(k_{2})e\gamma_{\mu}u(k_{1})\frac{-g^{\mu\nu}}{\tilde{p}^2}\bar{u}(p_{2})
  \gamma_{\nu} \gamma_{5}v(p_{3})\mathcal{F}_{\mathrm{NoR}}(\tilde{p}^2),
\end{equation} 
where $g_{\mathrm{NoR}}$ is an effective coupling constant determined by fitting to experimental data. The background form factor is chosen in an optimized form, compared with that used in Refs.~\cite{Xu:2015qqa,Wang:2017sxq},
\begin{equation}\label{}
  \mathcal{F}_{\mathrm{NoR}}(s)= \exp\biggl[-a\biggl(\sqrt{s}-\sum_fm_f\biggr)^2 -b\biggl(\sqrt{s}-\sum_fm_f\biggl)\biggr],
\end{equation}
where the $m_f$ represents the mass of the final states. The parameters $a$ and $b$ are fitted to experimental data.

Similarly, the amplitudes for the resonant contributions are written as
\begin{align}
    \mathcal{M}_{\rho(5/6S)} =& g_{\rho(5/6S)} \frac{eM_{\rho(5/6S)}^2}{f_{\rho(5/6S)}} \bar{v}(k_2)e\gamma_\mu u(k_1)\frac{-g^{\mu\nu}}{\tilde{p}^2} \nonumber\\
   &\times \frac{-g_{\nu\alpha}+\tilde{p}_{\nu}\tilde{p}_{\alpha}/\tilde{p}^2} {\tilde{p}^2-M_{\rho(5/6S)}^2+iM_{\rho(5/6S)}\Gamma_{\rho(5/6S)}} \bar{u}(p_2)\gamma^\alpha\nonumber\\
   &\times \gamma_{5}v(p_{3}), 
\end{align}
where $\tilde{p}=(k_1+k_2)$. 

In general, the interference effects from the contribution of different resonances should be taken into account. Thus, the total amplitude with relative phases can be written as 
\begin{align}\label{}
  \mathcal{M}_{e^+e^- \to p\bar{p}\pi^0}=& \mathcal{M}_{\mathrm{NoR}} +e^{i\phi_1}\mathcal{M}_{\rho(5S)} +e^{i\phi_2}\mathcal{M}_{\rho(6S)},
\end{align}
where the relative phases $\phi_1$ and $\phi_2$ are treated as free parameters fitted within the range $[0,2\pi]$. 

Then, the differential cross section for the process $e^+e^-\to \rho(5/6S) \to p\bar{p}\pi^0$ can be written as 
\begin{equation}\label{}
  d\sigma_{e^+e^-\to p\bar{p}\pi^0}=\frac{(2\pi)^4 |\overline{\mathcal{M}_{e^+e^- \to p\bar{p}\pi^0}}|^2}{4\sqrt{(k_1\cdot k_2)^2}}d\Phi_3,
\end{equation}
where the overline denotes averaging over the spins of the initial state and summing over the spins of the final state. The three-body phase-space element is given by
\begin{equation}\label{}
  d\Phi_3=\frac1{8(2\pi)^9\sqrt{s}}|\vec{p}_3^*||\vec{p}_2|d\Omega_3^*d\Omega_2dM_{\bar{p}\pi},
\end{equation}
where $\vec{p}_3^*$ and $\Omega_3^*$ denote, respectively, the three-momentum and solid angle of the antiproton in the rest frame of the $\bar{p}\pi$ subsystem. After integrating over the three-body phase space, the total cross section $\sigma_{e^+e^-\to p\bar{p}\pi^0}$ is obtained. 

\renewcommand{\arraystretch}{1.5}  
\begin{table}
  \centering
  \caption{The fitted parameters for the $e^+e^- \to p\bar{p}\pi^0$ process for two scenarios, Fit I and Fit II. }\label{fitpara}
  \setlength{\tabcolsep}{8pt}
  \begin{tabular}{l|lc}
    \toprule[1.50pt]
    \toprule[0.50pt]
    Parameters & Fit I & Fit II \\\hline
    $g_{\mathrm{NoR}}$ & 8.4$\pm$0.4 & 8.3$\pm$0.32 \\
    $a$ & -0.82$\pm$0.14 & -0.89$\pm$0.14 \\
    $b$ &  3.71$\pm$0.19 &  3.77$\pm$0.17 \\ 
    \hline
    $M_{\rho(5S)}$ [MeV] & 2400(Fixed) & 2422(Fixed) \\
    $\Gamma_{\rho(5S)}$ [MeV] & 150$\pm$70 & 153$\pm$22 \\
    $g_{\rho(5S)}$  & 21.9$\pm$3.4 & 25$\pm$10  \\
    $\phi_1$ [rad] & 3.54$\pm$0.12 & 3.82$\pm$0.09 \\ 
    \hline
    $M_{\rho(6S)}$ [MeV] & 2613(Fixed) & 2613(Fixed) \\
    $\Gamma_{\rho(6S)}$ [MeV] & 100$\pm$70 & 100$\pm$40 \\
    $g_{\rho(6S)}$ & 10$\pm$4 & 10.0$\pm$1.7 \\
    $\phi_2$ [rad] & 3.0$\pm$0.8 & 3.0$\pm$0.5 \\
    \hline
    $\chi^2/\text{d.o.f.}$ & 1.337 & 1.496 \\
    \toprule[0.50pt]\toprule[1.50pt]
    
  \end{tabular}
\end{table}


\begin{figure}[!htbp]
  \centering
  \begin{tabular}{l}
    \includegraphics[width=0.48\textwidth]{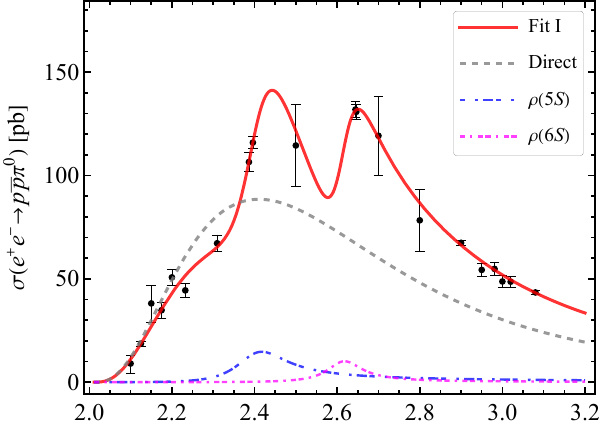} \\
    \includegraphics[width=0.48\textwidth]{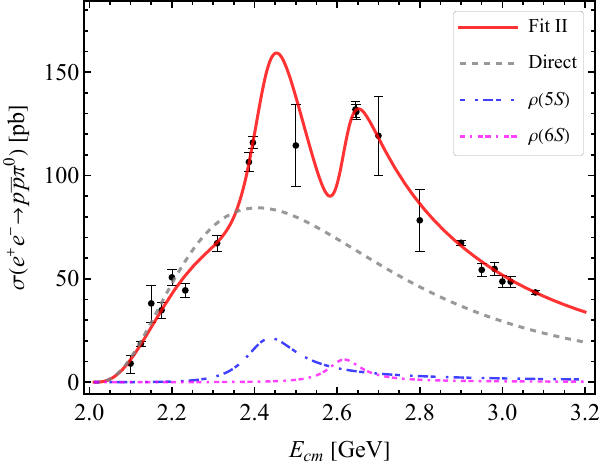} \\
  \end{tabular}
  \caption{The fitted Born cross sections for Fit I and Fit II scenarios of the $e^+e^- \to p\bar{p}\pi^0$ process \cite{BESIII:2024gjj} are shown by the red solid curves. The contributions from the direct background and the intermediate $\rho(5/6S)$ resonances are also displayed.}\label{plotfit}
\end{figure}

Based on the above formalism, we fit the $e^+e^- \to p\bar{p}\pi^0$ cross section data with nine free parameters. Owing to the limited statistics and the sparse energy coverage near the $\rho(5S)$ region, two fitting scenarios are considered: Fit~I, with the $\rho(5S)$ mass fixed at $2400\,\text{MeV}$, and Fit~II, with the $\rho(5S)$ mass fixed at $2422\,\text{MeV}$. Using the 20 experimental data points reported in Ref.~\cite{BESIII:2024gjj}, we obtain minimum $\chi^2/\mathrm{d.o.f.}$ values of 1.337 and 1.496 for Fits~I and II, respectively. The extracted resonance parameters, including total widths and coupling constants, are listed in Table~\ref{fitpara}. Although the fitted total widths of $\rho(5/6S)$ are somewhat larger than the sum for two-body decay widths estimated in the QPC model (Table~\ref{tab:branch}), they remain compatible within the present experimental uncertainties. From the fitted parameters, the branching fractions for $\rho(5/6S)\to p\bar{p}\pi^0$ are determined to be $\mathcal{B}[\rho(5S)] = 2.2\%$ (Fit~I), $\mathcal{B}[\rho(5S)] = 3.3\%$ (Fit~II), and $\mathcal{B}[\rho(6S)] = 2.4\%$, which do not contradict with usual expectations~\cite{ParticleDataGroup:2024cfk}. It can also be seen that if the $\rho$-resonance contribution were replaced by the $\omega$-resonance here, the resulting branching ratio for $\omega(5/6S) \to p\bar{p}\pi$ would be estimated at about $30\%$, which is obviously unphysical. This in turn supports the consistency of our initial treatment.

The fitted results, incorporating both the direct background and the $\rho(5/6S)$ resonance contributions, are displayed in Fig.~\ref{plotfit}, together with the BESIII experimental Born cross-section data \cite{BESIII:2024gjj}. As expected, the direct background term (Fig.~\ref{Feynman1}(d)) provides the dominant contribution and gives a smooth energy dependence without any resonance structure. However, it alone cannot reproduce the experimental data, which strongly supports the necessity of including the $\rho(5/6S)$ resonance contributions. In particular, the pronounced rise near 2.4 GeV provides compelling evidence for the presence of the anticipated $\rho(5S)$ state. 

Both Fit I and Fit II achieve good agreement with the measured cross sections. The main difference is that the $\rho(5S)$ peak in Fit II is slightly larger than that in Fit I. Because of the relatively larger $\rho(5S)$ contribution and smaller $\rho(6S)$ contribution in Fit II, the overall lineshape is characterized by one prominent peak followed by another smaller one. Although the precise $\rho(5S)$ lineshape is not well constrained due to the limited data quality, both fits require a clear Breit–Wigner structure, highlighting the role of an isovector resonance. Moreover, the total fitted curves show a slight shift of the resonance peaks toward higher energies, which arises naturally from interference effects. 

The interference plays an indispensable role in shaping the cross sections. On the one hand, destructive interference is observed around 2.3 GeV, where the total cross section even falls below the background term, and around 2.55 GeV, where a dip appears deeper than the standalone $\rho(6S)$ contribution. On the other hand, constructive interference occurs around 2.45 GeV and 2.65 GeV, where two pronounced peaks emerge and both are higher than the magnitudes of the individual resonances. These interference effects also slightly shift the effective peak positions away from the nominal $\rho(5/6S)$ masses. 

Unfortunately, the current experimental data suffer from large uncertainties and sparse energy bins. More precise and finely binned measurements in the future will be crucial for improving the reliability of the analysis and confirming the existence of these two long-term unestablished isovector light vector meson excitations.

\begin{figure}[tbph]
    \centering
\includegraphics[width=0.48\textwidth]{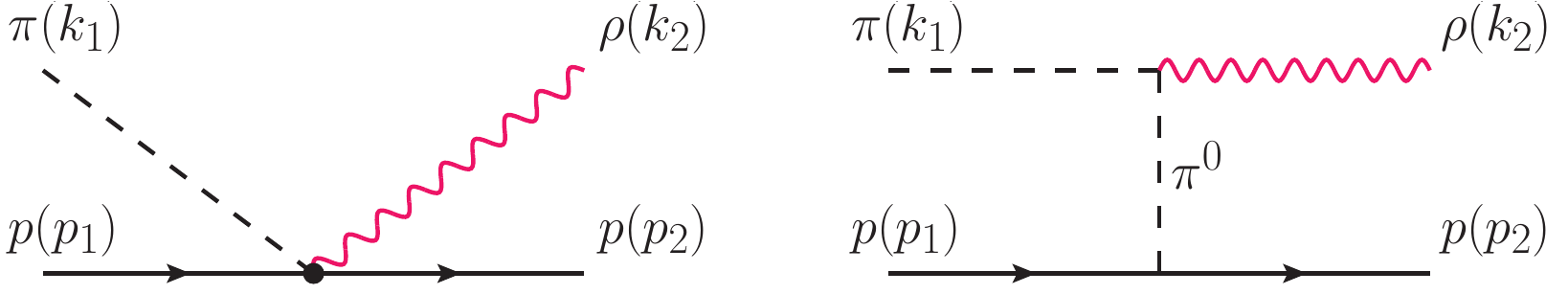}
    \caption{The contact and $t$-channel Feynman diagrams of the process $\pi p\to\rho(5/6S)p$. }
    \label{Feynman2}
\end{figure}

\section{production of $\rho(5/6S)$ resonances in $\pi p$ scattering}


Based on the fit procedure of the $e^+e^-\to p\bar{p}\pi^0$ reaction discussed in Sec.~\ref{sec:fit}, we can extract the effective contact couplings of the $\rho(5/6S)N\bar{N}\pi$ vertex, in which the results of Fit I are taken as input. With these couplings, we further investigate the production behaviors of $\rho(5/6S)$ in the $\pi p\to \rho(5/6S)p$ reaction, as illustrated in Fig.~\ref{Feynman2}. In addition, following the discussion in Sec.~\ref{sec:domi_anal}, we take into account the $t$-channel pion-exchange contribution, while neglecting the relatively suppressed $u$-channel nucleon-exchange diagram. 

For the $\pi NN$ interaction, we employ the well-established effective Lagrangian~\cite{Guo:2025mha},
\begin{equation}
    \mathcal{L}_{\pi NN}=\frac{g_{\pi NN}}{2M_N}\bar{N}\gamma^\mu\gamma_5\partial_\mu\vec{\pi}\cdot\vec{\tau} N,
\end{equation}
with the coupling constant fixed at $g_{\pi NN}=13.26$. 
For the other relevant vertex in the $t$-channel, we adopt the effective Lagrangian~\cite{Matsuyama:2006rp}
\begin{equation}
\mathcal{L}_{\rho\pi\pi} = g_{\rho\pi\pi}[\vec{\pi} \times \partial_\mu \vec{\pi}]\cdot \vec{\rho}^{\mu},
\end{equation}
where $\vec{\pi}$ and $\vec{\rho}^\mu$ denote the pion- and $\rho$-meson fields, respectively. The coupling constants $g_{\rho(5/6S)\pi\pi}$ are extracted from the partial decay branching ratios listed in Table~\ref{tab:branch} together with the fitted total widths in scheme I given in Table~\ref{fitpara}. 

The differential cross sections for the process $\pi(k_1) p(p_1)\to\rho(5/6S)(k_2)p(p_2)$ in the center-of-mass (CM) frame are expressed as
\begin{equation}\label{}
  \frac{\mathrm{d}\sigma_{\pi p\to\rho(5/6S)p}}{\mathrm{d}\cos\theta}=\frac1{32\pi s}\frac{|\vec{p}_2^{\mathrm{cm}}|}{|\vec{p}_1^{\mathrm{cm}}|}\overline{|\mathcal{M}_{\rho(5/6S)p}|^2},
\end{equation}
then integrating over $\cos\theta$ yields the production cross sections. 



\begin{figure}[tbph]
  \centering
  \includegraphics[width=0.48\textwidth]{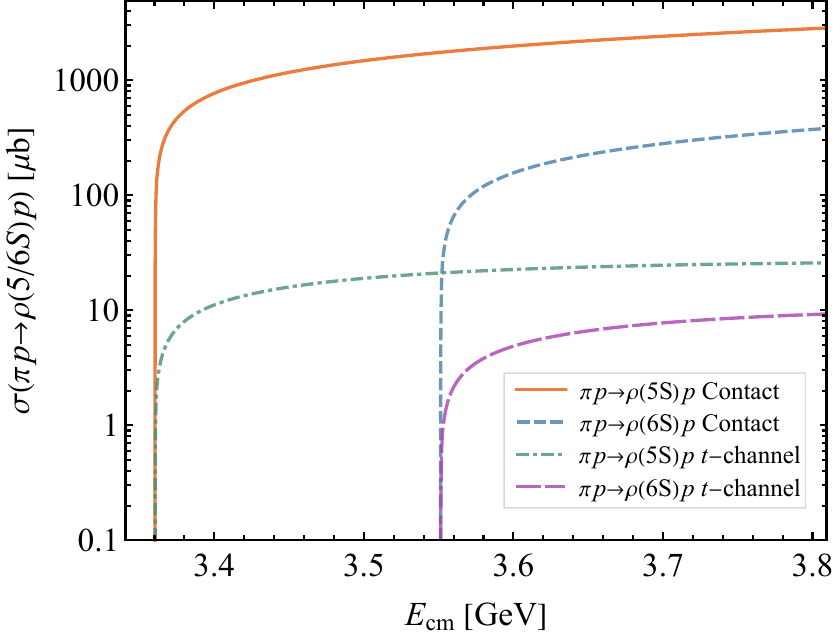}
  \caption{Production cross sections of the $p\bar{p}\to\pi^0\rho(5/6S)$ reaction, showing the contributions from the contact term and the $t$-channel exchange.}\label{fig:2-2tot}
\end{figure}

In Fig.~\ref{fig:2-2tot}, we present the production cross sections of the $\pi p\to\rho(5/6S)p$ process, including both the contact term and the $t$-channel pion-exchange contributions. The production behaviors of $\rho(5S)$ and $\rho(6S)$ are qualitatively similar: the cross sections increase steeply just above the corresponding thresholds, reflecting the opening of phase space, and subsequently level off as the center-of-mass energy increases. 

A clear hierarchy is observed between the two production mechanisms. For the contact term, the $\rho(5S)$ contribution exceeds that of the $\rho(6S)$ by approximately an order of magnitude, which can be attributed to the stronger effective coupling of the $\rho(5S)$ to the $p\bar{p}\pi^0$. In contrast, the $t$-channel pion-exchange contributions are significantly suppressed compared to the contact terms, being at least an order of magnitude smaller across the considered energy range. Interestingly, for the $t$-channel case, the peak cross section of $\rho(5S)$ is only about a factor of two larger than that of $\rho(6S)$, implying that in the considered energy region, the off-shell suppression of the pion propagator impacts both states in a similar way, such that the difference in effective couplings does not lead to as significant a hierarchy as observed in the contact contributions.

Overall, these results demonstrate that the contact interaction dominates the $\pi p\to\rho(5/6S)p$ production, while the $t$-channel exchange provides only a minor correction. Nevertheless, the $t$-channel contribution may still play a role in shaping the detailed energy dependence of the cross sections and could become relevant in precision studies. 

\renewcommand{\arraystretch}{1.5}  
\begin{table}[]
    \centering
    \caption{Branching cross sections for the reactions $\pi p\to (\rho(5/6S)\to M_{1}M_{2})p$, where the final states $M_{1}M_{2}$ are listed below. Cross sections are given in units of~$\mu\text{b}$. Centerdots (...) indicate channels for which the $\rho(5/6S)$ branching fractions are negligible.}
    \label{tab:cs}
    \setlength{\tabcolsep}{4pt} 
    \begin{tabular}{c|cccc|ccc}
        \toprule[1.50pt] \toprule[0.50pt]
            & \multicolumn{4}{c|}{$\rho(5S)$} & \multicolumn{3}{c}{$\rho(6S)$} \\ \hline
          $E_{cm}$ (GeV) & 3.4 & 3.5 & 3.6 & 3.8      & 3.6 & 3.7 & 3.8  \\ \hline
         $\sigma(\pi p\to \pi\pi p)$&  18&34&46&65 & 3&5&6 \\
         $\sigma(\pi p\to \pi\omega p)$&  12&22&30&42 & 1&2&2 \\
         $\sigma(\pi p\to \rho\rho(1450) p)$&  72&139&186&264 & 14&26&34 \\
         $\sigma(\pi p\to \pi\pi(1300) p)$&  65&124&166&236 & 11&21&27 \\
         $\sigma(\pi p\to \pi\omega(1420) p)$&  59&113&152&215 & 8&14&19 \\
         $\sigma(\pi p\to \pi a_2(1320) p)$&  32&61&81&115 & ...&...&... \\
         $\sigma(\pi p\to \omega a_0(1450) p)$&  60&116&155&220 & 6&11&15 \\
         $\sigma(\pi p\to \pi h_1(1170) p)$&  25&48&65&92 & ...&...&... \\
         \toprule[0.50pt] \toprule[1.50pt]
    \end{tabular}
\end{table}
\setlength{\extrarowheight}{0pt}

For reference in experimental analyses, Table~\ref{tab:cs} presents the three-body branching cross sections of the processes $\pi p \to (\rho(5/6S)\to M_{1}M_{2})p$ at several typical CM energies. The calculations are performed under the narrow-width approximation, which is well justified since the $\rho(5/6S)$ masses are much larger than their widths. The corresponding decay branching ratios of $\rho(5/6S)$ have been summarized in Table~\ref{tab:branch}. As shown in Table~\ref{tab:cs}, the branching cross sections increase with $E_{cm}$. Among the final states considered, $\rho\rho(1450)p$, $\pi\pi(1300)p$, $\pi\omega(1420)p$, and $\omega a_0(1450)p$ exhibit the largest measured cross sections. Nevertheless, the $\pi\pi p$ and $\pi\omega p$ final states are expected to be more favorable for experimental studies owing to their higher reconstruction efficiencies.

\section{Summary}

The spectroscopy of unflavored light vector mesons above 2.4 GeV remains experimentally unknown, despite its importance for understanding strong interaction dynamics and testing quark-model predictions of higher radial excitations. In this work, we have identified the process $e^+e^-\to p\bar{p}\pi^0$ as a particularly potential channel for exploring the missing $\rho(5S)$ and $\rho(6S)$ states. Owing to its relatively small continuum background and higher production threshold, this reaction offers a cleaner environment compared to the electron-positron annihilation to typical light-meson final states.  Our fits to the BESIII data of $e^+e^-\to p\bar{p}\pi^0$ within an effective Lagrangian approach shows that the inclusion of $\rho(5S)$ and $\rho(6S)$ is essential to reproduce the observed line shape, with extracted branching fractions to $p\bar{p}\pi^0$ below 4\%, consistent with usual expectations. The interference between the two $\rho$-meson resonances and the continuum background plays an important role, generating both constructive enhancements and destructive dips in the cross section distribution.


To provide complementary probes, we have further evaluated $\pi p \to \rho(5/6S)p$ production, including both contact and $t$-channel pion-exchange contributions. The results show that the contact term is dominant, while the $t$-channel provides only a minor correction. The predicted cross sections rise rapidly near threshold and reach the order of several tens to hundreds of $\mu$b at higher energies. Among various decay channels of $\rho(5/6S)$, $\rho\rho(1450)$, $\pi\pi(1300)$, $\omega a_0(1450)$ and $\pi\omega(1420)$ yield the largest branching cross sections, though the $\pi\pi p$ and $\pi\omega p$ modes remain the most favorable for experimental searches due to their reconstruction advantages.

Overall, our study indicates that the $e^+e^-\to p\bar{p}\pi^0$ and $\pi p \to \rho(5/6S)p$ processes provide  promising avenues for establishing the long-missing $\rho(5S)$ and $\rho(6S)$ states. Improved precision measurements with finer energy scans at BESIII, BelleII or future facilities, together with dedicated pion-beam experiments, will be crucial for confirming the existence of these excitations and deepening our understanding of the light vector meson spectroscopy above 2.4~GeV.

\section*{Acknowledgements}
D.~G. thanks Prof. Bing-Song Zou for useful discussion. 
D.~G. is supported by the National Natural Science Foundation of China under Grants No. 12347119, the China Postdoctoral Science Foundation under Grant No. 2023M740117, and the China National Postdoctoral Program for Innovative Talent No. BX20230379. 
J.-Z.~W. is supported by the National Natural Science Foundation of China under Grants No. 12405088 and the Start-up Funds of Chongqing University. 
Q.-S.~Z. is supported by the Natural Science Foundation of Inner Mongolia Autonomous Region (Grant No.2025QN01045), the Research Support Program for High-Level Talents at the Autonomous Region level in the Inner Mongolia Autonomous Region (Grant No. 13100-15112049), and the Research Startup Project of Inner Mongolia University (Grant No. 10000-23112101/101).


\begin{thebibliography}{100}

\bibitem{Klempt:2007cp}
E.~Klempt and A.~Zaitsev,
Glueballs, Hybrids, Multiquarks. Experimental facts versus QCD inspired concepts,
\href{https://doi.org/10.1016/j.physrep.2007.07.006}{Phys. Rept. \textbf{454}, 1-202 (2007)}.


\bibitem{Ochs:2013gi}
W.~Ochs,
The Status of Glueballs,
\href{https://doi.org/10.1088/0954-3899/40/4/043001}{J. Phys. G \textbf{40}, 043001 (2013)}.


\bibitem{BaBar:2014omp}
A.~J.~Bevan \textit{et al.} [BaBar and Belle],
The Physics of the B Factories,
\href{https://doi.org/10.1140/epjc/s10052-014-3026-9}{Eur. Phys. J. C \textbf{74}, 3026 (2014)}.


\bibitem{Belle-II:2018jsg}
E.~Kou \textit{et al.} [Belle-II],
The Belle II Physics Book,
\href{https://doi.org/10.1093/ptep/ptz106}{PTEP \textbf{2019}, no.12, 123C01 (2019)};
[erratum: \href{https://doi.org/10.1093/ptep/ptaa008}{PTEP \textbf{2020}, no.2, 029201 (2020)}].


\bibitem{Asner:2008nq}
D.~M.~Asner, T.~Barnes, J.~M.~Bian, I.~I.~Bigi, N.~Brambilla, I.~R.~Boyko, V.~Bytev, K.~T.~Chao, J.~Charles and H.~X.~Chen, \textit{et al.}
Physics at BES-III,
\href{https://ui.adsabs.harvard.edu/abs/2008arXiv0809.1869A/abstract}{Int. J. Mod. Phys. A \textbf{24}, S1-794 (2009)}.


\bibitem{BESIII:2020nme}
M.~Ablikim \textit{et al.} [BESIII],
Future Physics Programme of BESIII,
\href{https://doi.org/10.1088/1674-1137/44/4/040001}{Chin. Phys. C \textbf{44}, 040001 (2020)}.


\bibitem{Godfrey:1998pd}
S.~Godfrey and J.~Napolitano,
Light meson spectroscopy,
\href{https://doi.org/10.1103/RevModPhys.71.1411}{Rev. Mod. Phys. \textbf{71}, 1411-1462 (1999)}.



\bibitem{Fang:2020aqg}
S.~S.~Fang [BESIII],
Light Hadron Physics at BES Experiment, 
\href{https://doi.org/10.1142/9789811217739_0016}{Symposium on 30 Years of BES Physic, 116-122}.



\bibitem{Huang:2020dga}
G.~S.~Huang,
$R$ Measurements and QCD Studies at the BES Experiments, 
\href{https://doi.org/10.1142/9789811217739_0020}{Symposium on 30 Years of BES Physic, 145-151}.






\bibitem{ParticleDataGroup:2024cfk}
S.~Navas \textit{et al.} [Particle Data Group],
Review of particle physics,
\href{https://doi.org/10.1103/PhysRevD.110.030001}{Phys. Rev. D \textbf{110}, 030001 (2024)}.




\bibitem{BaBar:2006gsq}
B.~Aubert \textit{et al.} [BaBar],
Structure at 2175-MeV in $e^{+} e^{-} \to \phi f_0(980)$ Observed via Initial-State Radiation,
\href{https://doi.org/10.1103/PhysRevD.74.091103}{Phys. Rev. D \textbf{74}, 091103 (2006)}.



\bibitem{BaBar:2005hhc}
B.~Aubert \textit{et al.} [BaBar],
Observation of a broad structure in the $\pi^+ \pi^- J/\psi$ mass spectrum around 4.26~GeV/c$^2$,
\href{https://doi.org/10.1103/PhysRevLett.95.142001}{Phys. Rev. Lett. \textbf{95}, 142001 (2005)}.



\bibitem{Zhu:2005hp}
S.~L.~Zhu,
The Possible interpretations of $Y(4260)$,
\href{https://doi.org/10.1016/j.physletb.2005.08.068}{Phys. Lett. B \textbf{625}, 212 (2005)}.


\bibitem{Barnes:2002mu}
T.~Barnes, N.~Black and P.~R.~Page,
Strong decays of strange quarkonia,
\href{https:/doi.org/10.1103/PhysRevD.68.054014}{Phys. Rev. D \textbf{68}, 054014 (2003)}.


\bibitem{Pang:2019ttv}
C.~Q.~Pang,
Excited states of $\phi$ meson,
\href{https://doi.org/10.1103/PhysRevD.99.074015}{Phys. Rev. D \textbf{99}, 074015 (2019)}.


\bibitem{Wang:2012wa}
X.~Wang, Z.~F.~Sun, D.~Y.~Chen, X.~Liu and T.~Matsuki,
Non-strange partner of strangeonium-like state $Y(2175)$,
\href{https://doi.org/10.1103/PhysRevD.85.074024}{Phys. Rev. D \textbf{85}, 074024 (2012)}.


\bibitem{Ding:2007pc}
G.~J.~Ding and M.~L.~Yan,
$Y(2175)$: Distinguish Hybrid State from Higher Quarkonium,
\href{https://doi.org/10.1016/j.physletb.2007.10.020}{Phys. Lett. B \textbf{657}, 49-54 (2007)}.


\bibitem{Wang:2021gle}
J.~Z.~Wang, L.~M.~Wang, X.~Liu and T.~Matsuki,
Deciphering the light vector meson contribution to the cross sections of $e^+e^-$ annihilations into the open-strange channels through a combined analysis,
\href{https://doi.org/10.1103/PhysRevD.104.054045}{Phys. Rev. D \textbf{104}, 054045 (2021)}.


\bibitem{BESIII:2020xmw}
M.~Ablikim \textit{et al.} [BESIII],
Observation of a resonant structure in $e^{+}e^{-} \to \omega\eta$ and another in $e^{+}e^{-} \to \omega\pi^{0}$ at center-of-mass energies between 2.00 and 3.08 GeV,
\href{https://doi.org/10.1016/j.physletb.2020.136059}{Phys. Lett. B \textbf{813}, 136059 (2021)}.


\bibitem{BESIII:2021uni}
M.~Ablikim \textit{et al.} [BESIII],
Measurement of the $e^{+}e^{-}\rightarrow\omega\pi^{0}\pi^{0}$ cross section at center-of-mass energies from 2.0 to 3.08~GeV,
\href{https://doi.org/10.1103/PhysRevD.105.032005}{Phys. Rev. D \textbf{105}, 032005 (2022)}.


\bibitem{Zhou:2022wwk}
Q.~S.~Zhou, J.~Z.~Wang and X.~Liu,
Role of the $\omega(4S)$ and $\omega$(3D) states in mediating the $e^+e^-\to\omega\eta$ and $\omega\pi^0\pi^0$ processes,
\href{https://doi.org/10.1103/PhysRevD.106.034010}{Phys. Rev. D \textbf{106}, 034010 (2022)}.


\bibitem{Bai:2025knk}
Z.~Y.~Bai, Q.~S.~Zhou and X.~Liu,
Role of $4S-3D$ mixing in explaining the $\omega$-like $Y(2119)$ observed in $e^+e^-\to\rho\pi$ and $\rho(1450)\pi$,
\href{https://doi.org/10.1103/PhysRevD.111.054013}{Phys. Rev. D \textbf{111}, 054013 (2025)}.


\bibitem{BESIII:2024okl}
M.~Ablikim \textit{et al.} [BESIII],
Study of $e^+e^-\to\pi^+\pi^-\pi^0$ at $\sqrt{s}$ from 2.00 to 3.08~GeV at BESIII, 
\href{https://doi.org/10.1103/PhysRevD.110.032005}{Phys. Rev. D \textbf{110}, 032005 (2024)}.


\bibitem{Godfrey:1985xj}
S.~Godfrey and N.~Isgur,
Mesons in a Relativized Quark Model with Chromodynamics,
\href{https://doi.org/10.1103/PhysRevD.32.189}{Phys. Rev. D \textbf{32}, 189-231 (1985)}.


\bibitem{Barnes:1996ff}
T.~Barnes, F.~E.~Close, P.~R.~Page and E.~S.~Swanson,
Higher quarkonia,
\href{https://doi.org/10.1103/PhysRevD.55.4157}{Phys. Rev. D \textbf{55}, 4157-4188 (1997)}.


\bibitem{Ebert:2009ub}
D.~Ebert, R.~N.~Faustov and V.~O.~Galkin,
Mass spectra and Regge trajectories of light mesons in the relativistic quark model,
\href{https://doi.org/10.1103/PhysRevD.79.114029}{Phys. Rev. D \textbf{79}, 114029 (2009)}.


\bibitem{Li:2021qgz}
Z.~Y.~Li, D.~M.~Li, E.~Wang, W.~C.~Yan and Q.~T.~Song,
Assignments of the $Y(2040)$, $\rho(1900)$, and $\rho(2150)$ in the quark model,
\href{https://doi.org/10.1103/PhysRevD.104.034013}{Phys. Rev. D \textbf{104}, 034013 (2021)}.




\bibitem{He:2013ttg}
L.~P.~He, X.~Wang and X.~Liu,
Towards two-body strong decay behavior of higher $\rho$ and $\rho_3$ mesons,
\href{https://doi.org/10.1103/PhysRevD.88.034008}{Phys. Rev. D \textbf{88}, 034008 (2013)}.


\bibitem{Feng:2021igh}
J.~C.~Feng, X.~W.~Kang, Q.~F.~L{\"u} and F.~S.~Zhang,
Possible assignment of excited light S31 vector mesons,
\href{https://doi.org/10.1103/PhysRevD.104.054027}{Phys. Rev. D \textbf{104}, 054027 (2021)}.


\bibitem{Hilger:2015ora}
T.~Hilger, M.~Gomez-Rocha and A.~Krassnigg,
Light-quarkonium spectra and orbital-angular-momentum decomposition in a Bethe{\textendash}Salpeter-equation approach,
\href{https://doi.org/10.1140/epjc/s10052-017-5163-4}{Eur. Phys. J. C \textbf{77}, 625 (2017)}.



\bibitem{Branz:2010ub}
T.~Branz, T.~Gutsche, V.~E.~Lyubovitskij, I.~Schmidt and A.~Vega,
Light and heavy mesons in a soft-wall holographic approach,
\href{https://doi.org/10.1103/PhysRevD.82.074022}{Phys. Rev. D \textbf{82}, 074022 (2010)}.


\bibitem{Yu:2021ggd}
G.~L.~Yu, Z.~G.~Wang, X.~W.~Wang and H.~J.~Wang,
The ground states and the first radially excited states of D-wave vector {\ensuremath{\rho}} and {\ensuremath{\phi}} mesons,
\href{https://doi.org/10.1142/S0217751X21501979}{Int. J. Mod. Phys. A \textbf{36}, 2150197 (2021)}.



\bibitem{BESIII:2023sbq}
M.~Ablikim \textit{et al.} [BESIII],
Measurement of the cross sections for $e^+e^-\to\eta\pi^+\pi^-$ at center-of-mass energies between 2.00 and 3.08~GeV,
\href{https://doi.org/10.1103/PhysRevD.108.L111101}{Phys. Rev. D \textbf{108}, L111101 (2023)}.


\bibitem{BESIII:2020kpr}
M.~Ablikim \textit{et al.} [BESIII],
Measurement of the Born cross sections for $e^+e^- \to \eta^\prime \pi^{+}\pi^{-}$ at center-of-mass energies between $2.00$ and $3.08$~GeV,
\href{https://doi.org/10.1103/PhysRevD.103.072007}{Phys. Rev. D \textbf{103}, 072007 (2021)}.



\bibitem{Zhou:2025kmw}
Q.~S.~Zhou, Z.~Y.~Bai, H.~Xu, J.~Z.~Wang and X.~Liu,
Decoding the role of $\rho$ mesonic states for elucidating the $e^+e^-\to a_2(1320)\pi$ data and other reactions,
arXiv:2507.21554 [hep-ph].


\bibitem{Wang:2021abg}
L.~M.~Wang, S.~Q.~Luo and X.~Liu,
Light unflavored vector meson spectroscopy around the mass range of 2.4{\ensuremath{\sim}}3{\,}{\,}GeV and possible experimental evidence,
\href{https://doi.org/10.1103/PhysRevD.105.034011}{Phys. Rev. D \textbf{105}, 034011 (2022)}.



\bibitem{BESIII:2024gjj}
M.~Ablikim \textit{et al.} [BESIII],
Measurement of the $e^+e^-\to p\bar p\pi^0$ cross section at $\sqrt{s}=2.1000-3.0800$~GeV,
\href{https://doi.org/10.1103/PhysRevD.110.052006}{Phys. Rev. D \textbf{110}, 052006 (2024)}.


\bibitem{Achasov:2016zvn}
M.~N.~Achasov, A.~Y.~Barnyakov, K.~I.~Beloborodov, A.~V.~Berdyugin, D.~E.~Berkaev, A.~G.~Bogdanchikov, A.~A.~Botov, T.~V.~Dimova, V.~P.~Druzhinin and V.~B.~Golubev, \textit{et al.}
Updated measurement of the $e^+e^- \to \omega \pi^0 \to \pi^0\pi^0\gamma$ cross section with the SND detector,
\href{https://doi.org/10.1103/PhysRevD.94.112001}{Phys. Rev. D \textbf{94}, 112001 (2016)}.


\bibitem{BaBar:2007qju}
B.~Aubert \textit{et al.} [BaBar],
The $e^+e^-\to 2(\pi^+ \pi^-) \pi^0$, $2(\pi^+ \pi^-) \eta$, $K^+ K^-\pi^+\pi^-\pi^0$ and $K^+ K^-\pi^+ \pi^- \eta$ Cross Sections Measured with Initial-State Radiation, 
\href{https://doi.org/10.1103/PhysRevD.76.092005}{Phys. Rev. D \textbf{76}, 092005 (2007)};
[erratum: \href{ https://doi.org/10.1103/PhysRevD.77.119902}{Phys. Rev. D \textbf{77}, 119902 (2008)}].


\bibitem{BaBar:2022ahi}
J.~P.~Lees \textit{et al.} [BaBar],
Study of the reactions $e^+e^-\to K^+K^-\pi^0\pi^0\pi^0$, $e^+e^-\to K_S^0 K^\pm \pi^\mp \pi^0\pi^0$, and $e^+e^-\to K_S^0 K^\pm \pi^\mp\pi^+\pi^-$ at center-of-mass energies from threshold to 4.5~GeV using initial-state radiation,
\href{https://doi.org/10.1103/PhysRevD.107.072001}{Phys. Rev. D \textbf{107}, 072001 (2023)}.


\bibitem{Ketzer:2019wmd}
B.~Ketzer, B.~Grube and D.~Ryabchikov,
Light-Meson Spectroscopy with COMPASS,
\href{https://doi.org/10.1016/j.ppnp.2020.103755}{Prog. Part. Nucl. Phys. \textbf{113}, 103755 (2020)};
[erratum: \href{https://doi.org/10.1016/j.ppnp.2022.104000}{Prog. Part. Nucl. Phys. \textbf{128}, 104000 (2023)}]


\bibitem{Xiao:2019qhl}
L.~Y.~Xiao, X.~Z.~Weng, X.~H.~Zhong and S.~L.~Zhu,
A possible explanation of the threshold enhancement in the process $e^+e^-\rightarrow \Lambda\bar{\Lambda}$,
\href{https://doi.org/10.1088/1674-1137/43/11/113105}{Chin. Phys. C \textbf{43}, 113105 (2019)}.


\bibitem{Bai:2023dhc}
Z.~Y.~Bai, Q.~S.~Zhou and X.~Liu,
Higher strangeonium decays into light flavor baryon pairs like $\Lambda\bar{\Lambda}$, $\Sigma\bar{\Sigma}$, and $\Xi\bar{\Xi}$,
\href{https://doi.org/10.1103/PhysRevD.108.094036}{Phys. Rev. D \textbf{108}, 094036 (2023)}. 


\bibitem{Xu:2015qqa}
H.~Xu, J.~J.~Xie and X.~Liu,
Implication of the observed $e^{+}e^{-}\rightarrow p{\bar{p}}\pi ^0$ for studying the $p{\bar{p}}\rightarrow \psi (3770)\pi ^0$ process,
\href{https://doi.org/10.1140/epjc/s10052-016-4054-4}{Eur. Phys. J. C \textbf{76}, 192 (2016)}.


\bibitem{Wang:2017sxq}
J.~Z.~Wang, H.~Xu, J.~J.~Xie and X.~Liu,
Production of the charmoniumlike state Y(4220) through the $p\bar{p} \to Y(4220) \pi^0$ reaction,
\href{https://doi.org/10.1103/PhysRevD.96.094004}{Phys. Rev. D \textbf{96}, 094004 (2017)}.


\bibitem{Chen:2010nv}
D.~Y.~Chen, J.~He and X.~Liu,
Nonresonant explanation for the Y(4260) structure observed in the $e^+e^-\to J/\psi\pi^+\pi^-$ process,
\href{https://doi.org/10.1103/PhysRevD.83.054021}{Phys. Rev. D \textbf{83}, 054021 (2011)}.


\bibitem{Guo:2025mha}
D.~Guo, J.~Shi, I.~Strakovsky and B.~S.~Zou,
Analysis of $\Sigma^*$ via isospin selective reaction $K_Lp\to\pi^+\Sigma^0$,
\href{https://doi.org/10.1103/bm4l-4mmv}{Phys. Rev. D \textbf{112}, 034006 (2025)}.


\bibitem{Matsuyama:2006rp}
A.~Matsuyama, T.~Sato and T.~S.~H.~Lee,
Dynamical coupled-channel model of meson production reactions in the nucleon resonance region,
\href{https://doi.org/10.1016/j.physrep.2006.12.003}{Phys. Rept. \textbf{439}, 193-253 (2007)}.


\end{thebibliography}
\end{document}